\documentclass{gDEA2e}

\usepackage{epstopdf}
\usepackage{subfigure}
\usepackage{amsmath,mathrsfs}
\usepackage{amssymb, amsmath}
\usepackage{graphicx}
\usepackage{multirow}
\usepackage{tikz}
\usepackage{caption2}

\def\inte#1{
\displaystyle\mathop{#1\kern0pt}^\circ }

\def\virgp{\raise 2pt\hbox{,}}
\def\cdotpv{\raise 2pt\hbox{;}}

\def\eqdefa{\buildrel\hbox{\footnotesize def}\over =}

\def\C{\mathop{\mathbb C\kern 0pt}\nolimits}
\def\DD{\mathop{\mathbb D\kern 0pt}\nolimits}
\def\EE{\mathop{{\mathbb E \kern 0pt}}\nolimits}
\def\K{\mathop{\mathbb K\kern 0pt}\nolimits}
\def\N{\mathop{\mathbb N\kern 0pt}\nolimits}
\def\Q{\mathop{\mathbb Q\kern 0pt}\nolimits}
\def\R{\mathop{\mathbb R\kern 0pt}\nolimits}
\def\SS{\mathop{\mathbb S\kern 0pt}\nolimits}
\def\ZZ{\mathop{\mathbb Z\kern 0pt}\nolimits}
\def\TT{\mathop{\mathbb T\kern 0pt}\nolimits}
\def\P{\mathop{\mathbb P\kern 0pt}\nolimits}

\newcommand{\beq}{\begin{equation}}
\newcommand{\eeq}{\end{equation}}
\newcommand{\ben}{\begin{eqnarray}}
\newcommand{\een}{\end{eqnarray}}
\newcommand{\beno}{\begin{eqnarray*}}
\newcommand{\eeno}{\end{eqnarray*}}

\theoremstyle{plain}

\theoremstyle{definition}

\theoremstyle{remark}

\begin{document}


\articletype{}

\title{Neural networks for stock price prediction}

\author{
\name{Yue-Gang Song\textsuperscript{a},
Yu-Long Zhou\textsuperscript{b}$^{\ast}$,\thanks{$^\ast$Corresponding author. Email: yulong\_zhou@163.com}
Ren-Jie Han\textsuperscript{c}}
\affil{\textsuperscript{a}Business School, Henan Normal University, Xinxiang, P. R. China;
\textsuperscript{b}School of Mathematics, Yunnan Normal University, Kunming,  P. R.  China;
\textsuperscript{c}College of Economics, Sichuan University, Chengdu, P. R. China.
}
}

\maketitle

\begin{abstract}
Due to the extremely volatile nature of financial markets,
it is commonly accepted that stock price prediction is a task full of challenge. However in order to make profits or understand the essence of equity market, numerous market participants or researchers try to
forecast stock price using various statistical, econometric or even neural network models. In this work, we survey and compare the predictive power of five neural network models, namely, back propagation (BP) neural network, radial basis function (RBF) neural network, general regression neural network (GRNN), support vector machine regression (SVMR), least squares support vector machine regresssion (LS-SVMR). We apply the five models to make price prediction of three individual stocks, namely, Bank of China, Vanke A and Kweichou Moutai. Adopting mean square error and average absolute percentage error as criteria, we find BP neural network consistently and robustly outperforms the other four models.
\end{abstract}


\begin{classcode}62M10; 62M20; 91G70; 91G80; 91B84 \end{classcode}

\tableofcontents

\section{Introduction}

Price prediction in equity markets is of great practical and theoretical interest. On one hand, relatively accurate prediction brings maximum profit to investors. Many market participants, especially institutional ones, spend a lot of time and money to collect and analyze relevant information before making investment decision. On the other hand, researchers often use weather or not the price can be forecast to check market efficiency. Also, they invent, apply or adjust different models to improve the predictive power. Finding a 'good' method to more accurately forecast stock price will be a forever topic in both the academic field and the financial industry.
Equity price prediction is regarded as a challenging task in the financial time series prediction process since the stock market is essentially dynamic, nonlinear, complicated, nonparametric, and chaotic in nature \cite{abu96}. Besides, many macro-economical environments, such as political events, company¡¯s policies, general economic conditions, commodity price index, interest rates, investors¡¯ expectations, institutional investors¡¯ choices, and psychological factors of investors, are also the influencing factors \cite{tan07}.
In this paper, we apply five artificial intelligence (AI) models in the predicting research. Among the AI models, the back-propagation neural networks (BPNN), radial basis neural networks (RBFNN), general regression neural network (GRNN), support vector machine regression (SVMR), least squares support vector machine regression (LS-SVMR) are the most widely used and mature methods.
The back propagation neural network (BPNN) is successfully used in many fields, such as engineering \cite{sun18}, power forecasting \cite{mao13}, time series forecasting \cite{wan15}, stock index forecasting \cite{lu11}, stock price variation prediction \cite{hsi11}. BPNN is also useful in the economic field. \cite{wan11} proposed a hybrid forecasting model (Wavelet Denoising-based Back Propagation), which firstly decomposed the original data into multiple layers by wavelet transform, then a BPNN model was established by the low-frequency signal of each layer for predicting the Shanghai Composite Index (SCI) closing price.
The radial basis function neural network (RBFNN) is a feedforward neural network with a simple structure, which has a single hidden layer. \cite{sh15}applied RBFNN as a tool for nonlinear pattern recognition to correct the estimation error of the prediction of linear models in predicting two stock series in Shanghai and Shenzhen stock exchanges. In \cite{sun05}, the authors presented RBFNN¡¯s effectiveness in financial time series forecasting. RBFNN is proposed to overcome the main drawback of BPNN of easily falling into local minima in the training process. RBFNN have also been used in various forecasting areas and achieve good forecasting performance, with demonstrated advantages over BPNN in some applications \cite{lish10, kum10}.
The general regression neural network (GRNN), is put forward by Specht \cite{spe91}, shows its effectiveness in pattern recognition \cite{zha18}, stock price prediction \cite{luba13, pan10} and groundwater level prediction \cite{shi13}. \cite{pan10}showed the forecasting ability of GRNN in the prediction of closing stock price. However, their research is lack of comparison with other data mining models, which is also the limitation of other references in this paper.
Support Vector Machine (SVM), first developed by Vapnik \cite{vap00}, is based on statistical learning theory. Owing to its successful performance in classification tasks \cite{bur98, han17, ose97} and regression tasks \cite{smo04, vapgo96}, especially in time series prediction \cite{mull99} and financial related applications \cite{inc00}, it has drawn significant attention and thus earned intensive study. By using the structural risk minimization principle to turn the solving process into a convex quadratic programming problem, the support vector machine, obtains better generation performance and moreover the solution is unique and globally optimal \cite{lian16}.
 The least squares support vector machine regression (LS-SVMR) \cite{suy02}, based on structural risk minimization principle, is able to approximate any nonlinear system. As a reformulation of the SVM algorithm, LS-SVMR overcomes the drawbacks of local optima and overfitting in the traditional machine learning algorithm. In \cite{ism11}, the authors introduced a hybrid LSSVM model in time series forecasting. Reference \cite{heg14} proposed LSSVM in predicting S\&P 500 index price.
To our best knowledge, there is no article focused on comparing the effectiveness the above five algorithms we reviewed. In this study, we present this view by comparing the performance of the five neural networks, namely, BPNN, RBFNN, GRNN, SVM, LS-SVMR in predicting price of three individual stocks: Bank of China, Vanke A and Kweichou Moutai.

This paper is organized as follows. Section 2 is devoted to present the five neural network models.  In section 3, we describe the data, present the results, make deep analysis. We conclude in section 4.

\section{Methods}
In this section, we introduce the five neural network models.
\subsection{BP neural networks}

\begin{figure}[htbp]
\large
\begin{center}
\includegraphics[scale=1.0]{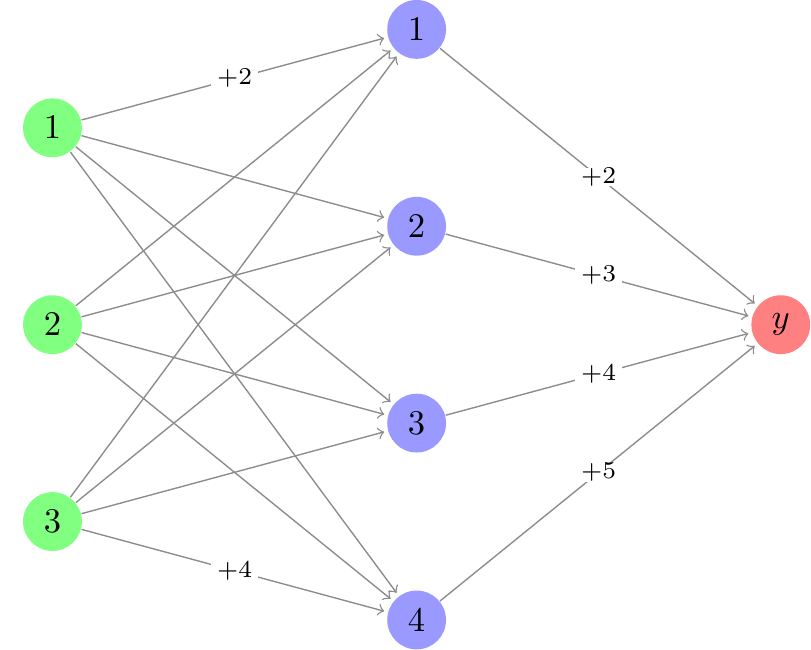}
\caption{Structure of a three layer neural network}
\label{NNthreeLayer}
\end{center}
\end{figure}

A neutral network generally contains one input layer, one or many hidden layers and one output layer. Suppose the total number of layers is $L$, we use $l$ to indicate a single layer. $l=1$ corresponds to the input layer, $l=2, \cdots, L-1$ corresponds to the hidden layers, and $l=L$ corresponds to the output layer. For example, figure \ref{NNthreeLayer} shows a neutral network containing only one hidden layer, which means $L=3$. Each layer contains one or many neurons, in figure \ref{NNthreeLayer}, the input layer contains 3 neurons, the hidden layer contains 4 neurons and the output layer contains only one neuron.

We use $w^{l}_{jk}$ to denote the weight for the connection from the $k$-th neuron in the $(l-1)$-th layer to the $j$-th neuron in the $l$-th layer. For illustration, we list some weights on the arrows in figure \ref{NNthreeLayer}. According to our notation, $w^{2}_{11}=2, w^{2}_{43}=4, w^{3}_{11}=2, w^{3}_{12}=3, w^{3}_{13}=4, w^{3}_{14}=5$. Explicitly, we use $b^{l}_{j}$ for the bias of the $j$th neuron in the $l$th layer. And we use $a^{l}_{j}$ for the activation of the $j$th neuron in the $l$th layer.
With these notations, the activation $a^{l}_{j}$ of the $j$th neuron in the $l$th layer is related to the activations in the $(l-1)$th layer by the equation
\ben \label{component-connection} a^{l}_{j}=\sigma(\sum_{k}w^{l}_{jk}a^{l-1}_{k}+b^{l}_{j}),\een
where the sigmoid function is defined as $\sigma(z)=\frac{1}{1-\exp(-z)}$.

To rewrite this expression in a matrix form we define a weight matrix $w^{l}$ for each layer $l$. The entries of the weight matrix $w^{l}$ are just the weights connecting to the $l$th layer of neurons, that is, the entry in the $j$th row and $k$th column is $w^{l}_{jk}$. Similarly, for each layer $l$ we define a bias vector $b^{l}$. The components of the bias vector are just the values $b^{l}_{j}$, one component for each neuron in the $l$th layer. And finally, we define an activation vector $a^{l}$ whose components are the activations $a^{l}_{j}$.

With these notations in mind, \eqref{component-connection} can be rewritten in the beautiful and compact vectorized form
\ben \label{matrix-connection} a^{l}=\sigma(w^{l}a^{l-1}+b^{l}).\een
Let $z^l$ be the weighted input to the neurons in layer $l$, that is
\beno z^l=w^{l}a^{l-1}+b^{l}.\eeno

The cost function is defined by the following quadratic form
\beno C=\frac{1}{2n}\sum_{x}||y(x)-a^{L}(x)||^{2} \eqdefa \frac{1}{n}\sum_{x}C_{x}.\eeno

We recall the Hadamard product $s \odot t$ between two vectors $s, t$ with the same length is defined by $(s \odot t)_{j}=s_{j}t_{j}$. The intermediate error function $\delta$ is computed as
\beno \delta^{L}=\nabla_{a}C\odot\sigma^{\prime}(z^{L}), ~~~~\delta^{l}=((w^{l+1})^{T}\delta^{l+1})\odot\sigma^{\prime}(z^{l}), l \geq 2.\eeno
\beno \frac{\partial C}{\partial b^{l}_{j}}=\delta^{l}_{j}, ~~~~\frac{\partial C}{\partial w^{l}_{jk}}=a^{l-1}_{k}\delta^{l}_{j}.\eeno
\beno \frac{\partial C}{\partial b^{l}}=\delta^{l}, ~~~~\frac{\partial C}{\partial w^{l}}=a^{l-1}\times\delta^{l}.\eeno
With learning rate $\eta$, the weights are learned by
\beno w^{l}_{jk}\longrightarrow w^{l}_{jk}-\eta\frac{\partial C}{\partial w^{l}_{jk}}, ~~~~b^{l}_{j}\longrightarrow b^{l}_{j}-\eta\frac{\partial C}{\partial b^{l}_{j}}.\eeno

Thanks to additivity of cost over sample, we can adopt the idea of stochastic gradient descent to speed up the learning.
To make these ideas more precise, stochastic gradient descent works by randomly picking out a small number $m$ of randomly chosen training inputs. We label those random training inputs $X_1, X_2, \cdots ,X_m$, and refer to them as a mini-batch. Provided the sample size $m$ is large enough we expect that the average value of the $\nabla C_{X_{j}}$ will be roughly equal to the average over all $\nabla C_{x}$, that is
\beno \frac{\sum_{j=1}^{m}\nabla C_{X_{j}}}{m}\approx\frac{\sum_{x}\nabla C_{x}}{n}=\nabla C.\eeno
Suppose $w_{k}$ and $b_l$ denote the weights and biases in our neural network. Then stochastic gradient descent works by picking out a randomly chosen mini-batch of training inputs, and training with those
\beno w_{k}\longrightarrow w^{\prime}_{k}=w_{k}-\frac{\eta}{m}\sum_{j=1}^{m}\frac{\partial C_{X_{j}}}{\partial w_{k}}.\eeno
\beno b_{l}\longrightarrow b^{\prime}_{l}=b_{l}-\frac{\eta}{m}\sum_{j=1}^{m}\frac{\partial C_{X_{j}}}{\partial b_{l}}.\eeno
Here the sums are over all the training examples $X_j$ in the current mini-batch.

\subsection{Radial basis function networks}
Radial basis function (RBF) networks typically have three layers: an input layer, a hidden layer with a non-linear RBF activation function and a linear output layer. Suppose the hidden layer has $I$ neurons, and the $i$th neuron centers at $c_{i}$ with its preferred value $w_{i}$.
The input can be modeled as a vector of real numbers $x \in \mathbb{R}^{n}$, the output of the network is then a scalar function of the input vector $\varphi: \mathbb{R}^{n}\rightarrow \mathbb{R}$, given by
\ben \label{output-RBF} \varphi(x)=\frac{\sum_{i=1}^{I}w_{i}\rho(||x-c_{i}||^{2})}{\sum_{i=1}^{I}\rho(||x-c_{i}||^{2})},\een
where $\rho(||x-c_{i}||^{2})=\exp(-\beta_{i}||x-c_{i}||^{2})$. Note that the output $\varphi(x)$ given input $x$ is the weighted average of $w_{i}$ with weight $\rho(||x-c_{i}||^{2})$.
The cost function is defined by the following quadratic form
\beno C=\frac{1}{2n}\sum_{x}||y(x)-\varphi(x)||^{2} \eqdefa \frac{1}{n}\sum_{x}C_{x}.\eeno
The parameters $w_{i}, c_{i}, \beta_{i}$ are selected by decreasing $C$.

\subsection{General regression neural networks}
Generalized regression neural network (GRNN) essentially belongs to radial basis neural networks. GRNN was suggested by D.F. Specht in 1991.
Recall the framework of RBF network, the number of neurons in the hidden layer is the same as the sample size $n$ of training data. Moreover, the center of the $i$th neuron is just the $i$th sample $x_{i}$, and the preferred value $w_{i}$ is set to be the desired output $y_{i}=y(x_{i})$. Then the output of a new input $x$ is
\ben \label{direct-output-GRNN} \varphi(x)=\frac{\sum_{i=1}^{I}y_{i}\rho(||x-x_{i}||^{2})}{\sum_{i=1}^{I}\rho(||x-x_{i}||^{2})},\een
where $\rho(||x-c_{i}||^{2})=\exp(-\beta||x-c_{i}||^{2})$.
Note that the output given input $x$ is just the weighted average of $y_{i}$ with weight $\rho(||x-x_{i}||^{2})$.
We remark that GRNN directly produces a predicted value without training process. One only needs to select a suitable smoothing parameter $\beta$ to implement GRNN.

If the number of neurons in the hidden layer stayed at the sample size $n$ of training data on each prediction, we call it a static GRNN. Instead, as new observations come, we may increase the number of neurons. We call such a pattern as a dynamic GRNN. We choose a dynamic GRNN in the current paper since it has stronger prediction power. A dynamic GRNN has long memory and updates timely.

\subsection{Support vector regression}
A version of SVM for regression (SVR) was proposed in 1996 by Vladimir N. Vapnik et al. The model aims to find a linear function of $x$ to predict $y$, namely,
\ben \label{output-SVR} f(x)=w \cdot x + b.\een
Let $\epsilon$ be error acceptance. Then one wants to find optimal $w$ and $b$ such that
\beno \sum_{i}^{n}|f(x_{i})-y_{i}|\leq \epsilon.\eeno
A modified problem is to solve
\beno \min \frac{1}{2}||w||^{2},~~~~s.t. ||Xw+bI_{n}-Y|| \leq \epsilon.\eeno
Here $Y=(y_{1},\cdots,y_{n})^{T}$ and $X$ is a matrix with its transpose $X^{T}=(x_{1},x_{2},\cdots,x_{n})$, the $i$th column of which is just $x_{i}$.

The linear predictor \eqref{output-SVR} cannot reveals possible nonlinear relation between $x$ and $y$, thus manifests its limitation. For a general (thus may be nonlinear) function $\phi: \mathbb{R}^{m} \rightarrow \mathbb{R}^{m}$, with $m$ being the dimension of $x$, one may adopt the following predictor
\ben \label{nonlinear-output-SVR} f(x)=w^{T}\phi(x)+b.\een
Different function $\phi$ corresponds to different kernel function in application. See next subsection for more kernel functions.

\subsection{Least squares SVM}
For some function $\phi: \mathbb{R}^{m} \rightarrow \mathbb{R}^{m}$, suppose we instead use $f(x) = w^{T}\phi(x)+b$ to predict $y(x)$.
Replace the inequality constraint by equality constraint, the Least squares SVMR reads
\beno \min \frac{1}{2}||w||^{2}+\frac{\gamma}{2}||\xi||^{2},~~~~s.t. ~~~~\phi(X)w+bI_{n}-Y=\xi.\eeno
Here $\phi(X)=(\phi(x_{1}),\cdots,\phi(x_{n}))^{T}$. Compared to SVM, LS-SVM requires less computation cost.

The solution of LS-SVM regressor will be obtained after we construct the Lagrangian function:
\beno L(w,b,\xi,\lambda)=\frac{1}{2}||w||^{2}+\frac{\gamma}{2}||\xi||^{2}+\lambda^{T}(\phi(X)w+bI_{n}-Y-\xi).\eeno
Take derivatives w.r.t. $w,b,\xi,\lambda$, and set the resulting functions to be zero, one has
\begin{equation}\label{take-derivatives} \left\{ \begin{aligned}
&\frac{\partial L}{\partial w}=0 ~~~~\Rightarrow w=\phi(X)^{T}\lambda=\sum_{i}^{n}\lambda_{i}\phi(x_{i});\\
&\frac{\partial L}{\partial b}=0 ~~~~\Rightarrow I_{n}^{T}\lambda=\sum_{i}^{n}\lambda_{i}=0;\\
&\frac{\partial L}{\partial \xi}=0 ~~~~\Rightarrow \lambda=\gamma\xi;\\
&\frac{\partial L}{\partial \lambda}=0 ~~~~\Rightarrow \phi(X)w+bI_{n}-y=\xi.
\end{aligned} \right.
\end{equation}
Rearrange the above system of equations after eliminating $w, \xi$, we have
\begin{equation}       
\left(\begin{array}{cc}   
0 & I^{T}_{n} \\  
I_{n} & \Omega+\gamma^{-1}I_{nn}   
\end{array}\right) \left(\begin{array}{c}   
b  \\  
\lambda  
\end{array}\right)=    \left(\begin{array}{c}   
 0  \\  
 Y
\end{array}\right),             
\end{equation}
where $I_{nn}$ is the $n\times n$ identity matrix and $\Omega$ is a $n\times n$ matrix defined by $\Omega_{ij}=\phi(x_{i})^{T}\phi(x_{j})=K(x_{i},x_{j})$. Once the previous equation is solved, the predicted value $f(x)$ for input $x$ is given by
\beno f(x)=\lambda^{T}\phi(X)\phi(x)+b=\sum_{i}^{n}\lambda_{i}K(x_{i},x)+b.\eeno The kernel function $K(x_{i}, x_{j})$ has many forms:
\begin{itemize}
\item Linear kernel: $K(x_{i}, x_{j})=x_{i}^{T}x_{j}$,
\item Polynomial kernel of degree $d$: $K(x_{i}, x_{j})=({1+x_{i}^{T}x_{j}/c})^{d}$,
\item RBF kernel : $ K(x_{i}, x_{j})=\exp ({-\|{x_{i}-x_{j}}\|^{2}/\sigma ^{2}})$,
\item MLP kernel : $ K(x_{i}, x_{j})=\tanh({k x_{i}^{T}x_{j}+\theta})$,
\end{itemize}
where $d, c, \sigma, k, \theta$ are constants. We remark that the linear kernel corresponds to the linear function $\phi(x)=x$. The most commonly used kernel is the RBF kernel.

\section{Data, Results and Discussions}
In this section, we describe the data we used and then present the results. Also we make some deep analysis of the results. In addition, we show the BP algorithm is stable and the market is inefficient.
\subsection{Data description}
In this work, we study the weekly adjusted close price of three individual stocks: Bank of China(601988), Vanke A(000002) and Kweichou Moutai(600519). Each price data has a sample size of 427, ranging from 3-January-2006 to 11-March-2018. As usual, we split the whole data set into a training set (80\%) and a test set (20\%).

\begin{table}[!htbp]
\centering
\caption{Price Range}\label{PriceRange}
\begin{tabular}{cccc}
\hline
Name&Bank of China& Vanke A& Kweichou Moutai\\
\hline
Lowest price&2.00& 5.65&81.13\\
Highest Price&5.01& 40.04&788.42\\
\hline
\end{tabular}
\end{table}

We intentionally select the three stocks based on such an observation: they are totally different in price scale. As showed in Table \ref{PriceRange}, the price of Bank of China is about within 2-5RMB,  Vanke A(000002) is approximately in the range 5-40RMB, Kweichou Moutai has a wide range 80-800RMB. Actually, Kweichou Moutai ranks first in terms of price per share among all stocks listed in the only two stock exchanges of Mainland of China: Shanghai and Shenzhen.

Let us use $\{S_{i}\}_{1 \leq i \leq 427}$ to denote the time series of price. We use three previous periods to predict the price of next period. More precisely, we set $x_{i}=(S_{i},S_{i+1},S_{i+2})$ and $y_{i}=S_{i+3}$ for $1 \leq i \leq 424$. Then we regard $(x_{i}, y_{i})$ as one sample. That is,  for an input $x_{i}$, its desired output is $y_{i}$. Note that we use weekly data, thus the information contained in the price is assumed to be effective within one month.

\subsection{Hyper-Parameters}

We adopt a neural network with three layers, which contains only one hidden layer. The input layer has three neurons, the output layer has only one neuron which represents the predicted value. To determine the number of neurons in the hidden layer, by rule of thumb, we apply the following formula
\beno m = \sqrt{0.43 l n +0.12l^{2}+2.54n+0.77l+0.35}+0.51,\eeno
where $l$ is the number of neurons in the output layer, and $n$ is the number of neurons in the input layer. With $l=1, n=3$, we get $m=3$ after rounding to integer. The learning rate $\eta=0.01$ is chosen after amounts of tests.

On the implementation of RBF, SVMR, LS-SVMR, we use standard R packages. When applying GRNN,  we choose $\beta=20, 0.5, 0.0005$ for Bank of China, Vanke A, Kweichou Moutai respectively, which reflecting the different price scales of the three stocks.

\subsection{Results}
Table \ref{ResultComparison} shows the performance of the five neural network models. From which, we can see all the five models have some predictive power. Even the worst one, GRNN, has MAPE not exceeding $5\%$, which is very satisfactory considering we are forecasting stock price rather than volatility.

Across all the three stocks, in terms of both MSE and MAPE, BP neural network outperforms the other four models. One may refer to Figure \ref{ForecstBC} in the next subsection for a more intuitive view of the accuracy of prediction for Bank of China with BP method.
SVMR ranks second consistently across the three stocks. However, in terms of both MSE and MAPE, results from SVMR are greater than that of BP by at least $10\%$. Moerover, on the prediction of Bank of China and Vanke A, BP surpasses SVMR by at least $20\%$ under both criteria.

\begin{table}[!htbp]
\centering
\caption{Results of the five methods}\label{ResultComparison}
\begin{tabular}{c|c|ccccc}
\hline
\multicolumn{2}{c|}{Method}&BP&RBF&GRNN&SVMR&LS-SVMR\\
\hline
\multirow{2}*{Bank of China}&MSE&0.009&0.014&0.020&0.012&0.018\\
&MAPE&0.019&0.025&0.024&0.023&0.028\\
\hline
\multirow{2}*{Vanke A}&MSE&2.976&4.686&6.036&3.422&5.472\\
&MAPE&0.049&0.065&0.067&0.059&0.072\\
\hline
\multirow{2}*{Kweichou Moutai}&MSE&395.1&740.1&1103.6&407.4&405.5\\
&MAPE&0.026&0.036&0.048&0.029&0.027\\
\hline

\end{tabular}
\end{table}

We can not tell which one of RBF and LS-SVMR is better. As showed in table \ref{ResultComparison}, on the prediction of Bank of China and Vanke A, RBF is more accurate than LS-SVMR, while on the prediction of Kweichou Moutai, LS-SVMR has a better performance.
Overall, they share a similar accuracy level of prediction. At last, GRNN behaves worst consistently across the three stocks.

One reason we could guess for the best performance of BP over other methods  is that the latter four models all involve the mostly used kernel function: $\exp(-|x|^{2})$. To check this guess, we use other kernels to apply the second best model: SVMR. Table \ref{KernelComparison} gives the results for the above mentioned four kernels.
\begin{table}[!htbp]
\centering
\caption{Comparison of four kernels in SVMR}\label{KernelComparison}
\begin{tabular}{c|c|cccc}
\hline
\multicolumn{2}{c|}{Kernel}&Linear&Polynomial&Sigmoid&RBF\\
\hline
\multirow{2}*{Bank of China}&MSE&0.010&0.010&0.011&0.012\\
&MAPE&0.019&0.020&0.021&0.023\\
\hline
\multirow{2}*{Vanke A}&MSE&2.993&3.292&3.515&3.422\\
&MAPE&0.050&0.053&0.055&0.059\\
\hline
\multirow{2}*{Kweichou Moutai}&MSE&395.6&403.5&405.7&407.4\\
&MAPE&0.027&0.028&0.028&0.029\\
\hline
\end{tabular}
\end{table}

We have two remarks on Table \ref{KernelComparison}. On one hand, linear kernel is the best in this prediction task, and consistently outperforms the other three kernels. Although RBF kernel is the default kernel in many package due its flexibility to various data resources, it is not good enough here. Thus we should try other kernels to make comparison when doing similar predicting projects.
On the other hand, BP stills beats SVMR with linear kernel, even if the advantage is not obvious now. They share a similar prediction error maybe thanks to the fact they both involve weighted average, which capturing some linear relation in the network.

\subsection{Two more discussions: stability of BP and market inefficiency}

When implementing BP algorithm, one needs to initialize the weights randomly, which causes instability of the result.
To show that the result of BP is stable, we train the neural network for 100 times, and compute its mean and standard deviation. Table \ref{Stability} helps us to eradiate this concern since the standard deviations are extremely small compared to the scale of their corresponding mean values. That is, the result of every experiment is reliable.

\begin{table}[!htbp]
\centering
\caption{100 times experiment}\label{Stability}
\begin{tabular}{c|c|cc}
\hline
\multicolumn{2}{c|}{Statistics}&Mean&Std.\\
\hline
\multirow{2}*{Bank of China}&MSE&0.009&$4.8 \times 10^{-5}$\\
&MAPE&0.019&0.0001\\
\hline
\multirow{2}*{Vanke A}&MSE&2.976&0.0067\\
&MAPE&0.049&0.0001\\
\hline
\multirow{2}*{Kweichou Moutai}&MSE&395.1&1.3728\\
&MAPE&0.026&$7.7 \times 10^{-5}$\\
\hline
\end{tabular}
\end{table}

\begin{figure}[htbp]
\large
\centering
\includegraphics[scale=0.7]{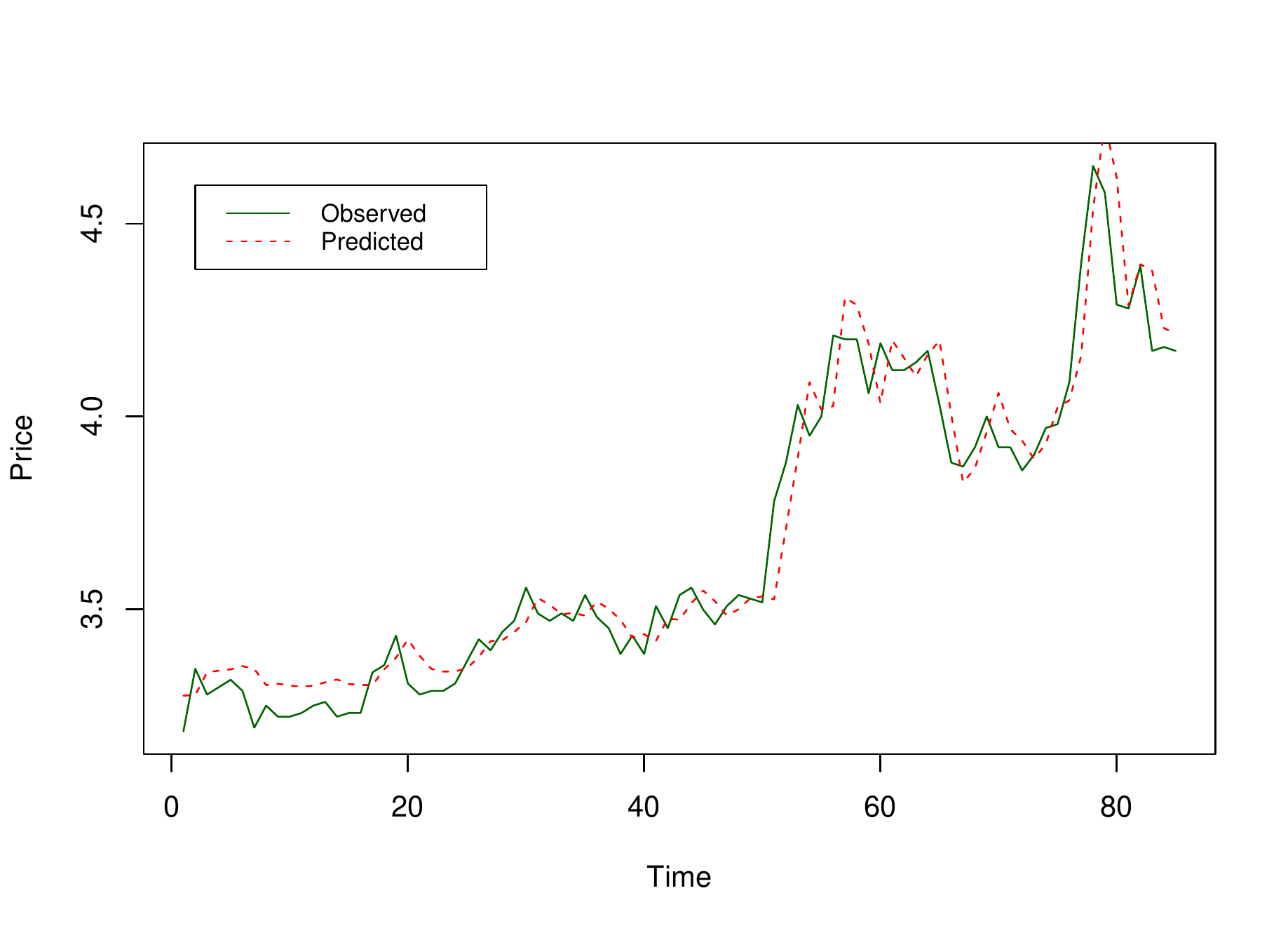}
\caption{Forecast of Bank of China}
\label{ForecstBC}
\end{figure}

Figure \ref{ForecstBC} plots the observed and predicted prices of Bank of China. It can be seen clearly that the predicted values fit well the observed ones. Also the turning points are forecasted quite timely. When there is a trend in the actual price, the predicted value follows accordingly and closely.

At a first glance, the network needs at least one period to react or assimilate new information. Actually, it is a false appearance that the predicted values lag one period of the observed values. More precisely, suppose $y_{t}$ is the observed price, $\hat{y}_{t}$ the predicted price, then it seems $y_{t}\approx\hat{y}_{t+1}$ from figure \ref{ForecstBC}, which means the best predicted value is just the price of previous period. Alternatively, the stock price process is Markovian.
If such a phenomenon is true, then the market is efficient and thus unpredictable.

\begin{figure}[htbp]
\large
\centering
\includegraphics[scale=0.7]{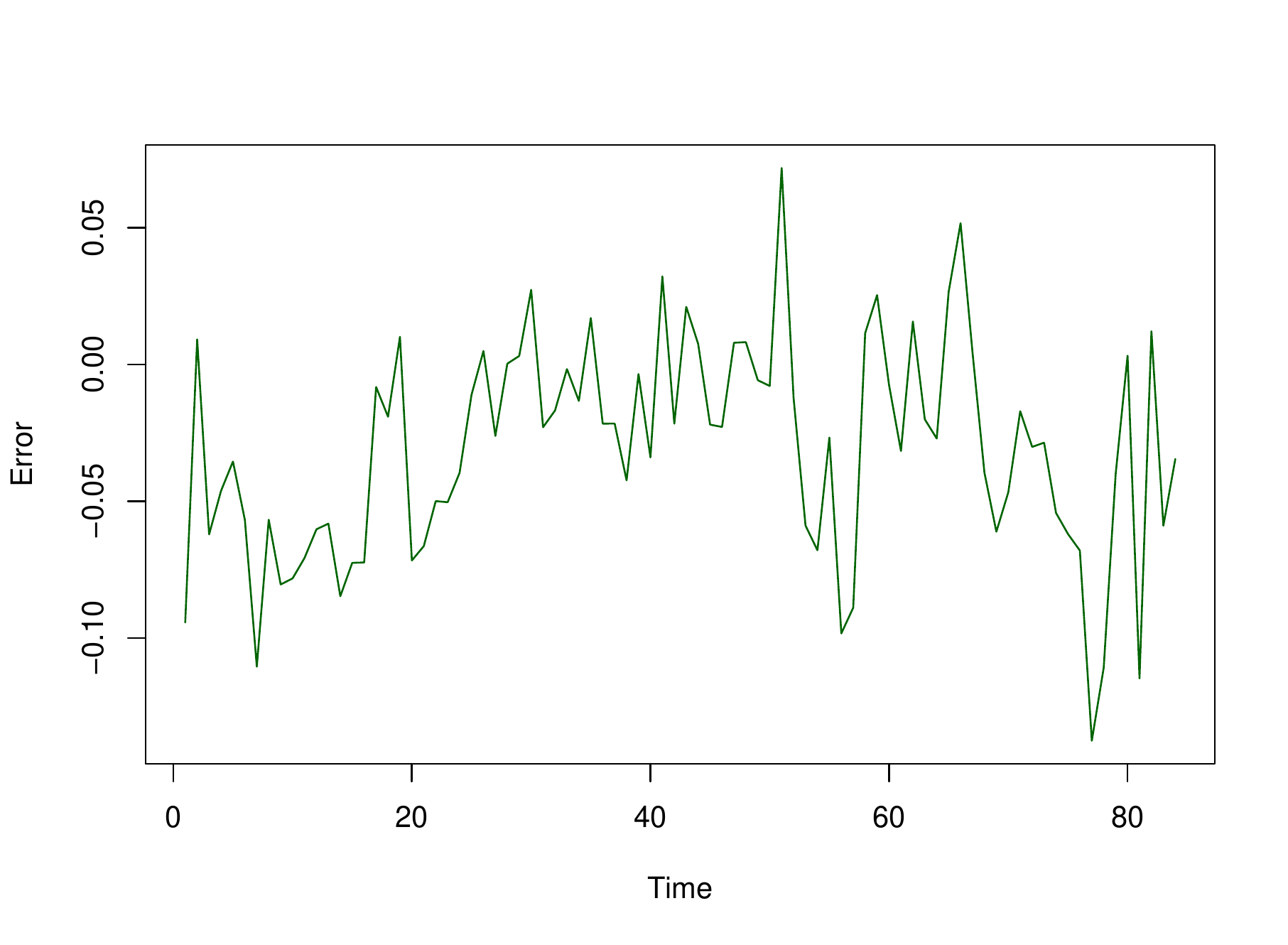}
\caption{Lag One Error}
\label{ErrorLagOne}
\end{figure}

To prove the market is actually inefficient, we take difference $e_{t}=y_{t}-\hat{y}_{t+1}$ and plot the difference series $e$ in figure \ref{ErrorLagOne}. Note that in figure \ref{ErrorLagOne}, the error is not centered at 0. Actually it has a bias towards to negative values. That is, on average, $y_{t}<\hat{y}_{t+1}$, which contradicts market efficiency.

\section{Conclusion}
In this work, we have successfully demonstrated that the five neural network models are all able to effectively extract meaningful information from past price. With evidence from forecast accuracy of three unrelated stocks, we find BP beats other four models consistently and robustly.
Also, by implementing the algorithm many times and checking the standard deviation, the stability of BP is observed and confirmed. Based on our trial on different kernels, we advise readers of the current paper not take the default kernel for granted and 'descrisized'  other kernels.
For our own interest, we test the error series and destroy the market efficiency hypothesis. In our future research, we will investigate other more involved neural networks to complete the current tentative work.


\end{document}